\documentstyle[12pt]{article}
\begin{document}
\setlength{\baselineskip}{.7cm}
\renewcommand{\thefootnote}{\fnsymbol{footnote}}
\sloppy

\newcommand \be  {\begin{equation}}
\newcommand \bea {\begin{eqnarray} \nonumber }
\newcommand \ee  {\end{equation}}
\newcommand \eea {\end{eqnarray}}

\begin{center}
\centering{\bf \Large Large deviations and portfolio optimization}
\end{center}

\begin{center}
\centering{Didier Sornette$^{1,2}$\\
{\it $^1$ Department of Earth and Space Science\\ and Institute of Geophysics
and Planetary Physics\\ University of California, Los Angeles,
California 90095\\
$^2$ Laboratoire de
Physique de la Mati\`ere Condens\'ee, CNRS UMR6622\\ Universit\'e des
Sciences, B.P. 70, Parc Valrose, 06108 Nice Cedex 2, France
}
}
\end{center}

\date{\today}

\vskip 2cm

{\bf Abstract}\,: Risk control and optimal diversification constitute a major focus
in the finance and insurance industries as well as, 
more or less consciously, in our everyday life. We present a
discussion of the characterization of risks and of the optimization of 
portfolios that starts from a simple illustrative model and ends by a general functional
integral formulation. A major theme is that risk, usually thought one-dimensional 
in the conventional mean-variance approach, has to be addressed by the full distribution 
of losses. Furthermore, the time-horizon of the investment is shown to play a major role.
We show the importance of accounting for large fluctuations and use the theory of Cram\'er for
large deviations in this context.
We first treat a simple model with a single risky asset that examplifies 
 the distinction between the average return and the typical return, the 
role of large deviations in multiplicative processes, and the 
different optimal strategies for the investors depending on their size. 
We then analyze the case of assets whose price variations are distributed according to 
exponential laws, a situation that is found to describe reasonably well daily price variations.
Several portfolio optimization strategies are presented that aim at
controlling large risks. We end by extending the standard mean-variance portfolio
optimization theory, first within the quasi-Gaussian approximation and then 
using a general formulation for non-Gaussian correlated assets in terms of 
the formalism of functional integrals developed in the field theory of critical phenomena.

\pagebreak

\section{Introduction}

Finance and insurance are all about risk, and risk is usually thought to
be all encapsulated by variances and covariances. In practice however, variances
and covariances are quite unstable with the existence of
intermittent bursts of high volatilities. More generally, risk
is embedded in the distribution of asset returns and not solely in variance. 
In finance and  in fact in many of our actions, we have to ponder the choice between risky
alternatives with different rewards. How to make the best choice(s)?

``Don't put all your eggs in one basket'' is a familiar adage. 
The idea is then to try to minimize
somehow the risk by diversification, a procedure that
is exemplified by the problem of portfolio selection and optimization in finance\,:
how to choose a basket of assets which maximizes your profit and minimizes your risk?

While these considerations seem quite straightforward and reflect the common sense, 
putting them in practice provides some surprise. Indeed, the very problem of
the quantification of the notion of risk does not find a simple general unambiguous answer. 
Even if the stochastic process 
describing the set of potential
profits and losses is stationary and can thus be described by a well-defined probability
distribution, the problem of ordering distributions does not have a unique answer.
This means that, in general, there does not exist a unique universal measure of risk, 
a single number that can be attributed to a risky situation. This is in contrast to the 
much better known problem of ordering numbers. As soon as one has to deal with 
pairs of numbers, vectors, matrices,
and even more so distributions, the problem of ordering is much harder
and in many cases has not a unique answer. There is a vast litterature that addresses this question
dealing with utility functions \cite{LevyM,Kroll}, stochastic dominance and so on
\cite{HuangLit}.

In section 2, we first present a simple model of a portfolio made of one risky and one riskless asset, which 
allows us to introduce the key ideas, namely the importance of the time-horizon of 
the investment, the distinction between the average return and the typical return, the 
role of large deviations in multiplicative processes, and the 
different optimal strategies for the investors depending on their size. In section 3, 
we analyze the case of assets whose price variations are distributed according to 
exponential laws, a situation that is found to describe reasonably well daily price variations.
This parametric case is used to illustrate the portfolio optimization strategies that aim
to control large risks. In section 4, we summarize first the standard mean-variance portfolio
optimization theory. We then show how to extend it within the quasi-Gaussian approximation. 
Finally, we provide a general formulation for non-Gaussian correlated assets in terms of 
the formalism of functional integrals developed in the field theory of critical phenomena.

\section{An exactly soluble case\,: one risky and one riskless assets}

Let us assume that the financial world is only made of a riskless asset and a risky asset.
Without loss of generality, the return of the riskless asset can be taken zero by a suitable
change of frame. Time is discrete in units of time $\tau$ taken to be unity. The return
over a unit time period $\tau$ of the risky asset is a random variable
$-\infty < r < +\infty$ without correlations, taken from a distribution $P(r)$. 
An investor has a wealth equal to unity at $t=0$. Should he invest a fraction $f$ 
in the risky asset? What is the optimal fraction $f$?

We start by treating this simple case because it is particularly useful to introduce
the key concepts that will be useful for the general problem of the diversification and 
optimization of portfolios. The standard mean-variance approach comes about 
very naturally and this models allows us to discuss its limits and generalization, in particular
to take into account the existence of large deviations.

\subsection{General case}

Here, we assume that the problem is ``separable'', i.e. $f$ is a constant 
independently of the wealth. A more general case consists in evolving $f$ as a function 
of the wealth of the investor, to account for the phenomenon that the degree of risk aversion
is a function of wealth. This is usually addressed using the formalism of utility functions, which
is mathematical device that quantifies the degree of satisfaction of an investor. Within this 
approach, the optimal fraction becomes a function $f(S(t), t)$ of time and of the wealth $S(t)$.

The choice $f=0$ corresponds to an investor who is absolutely opposed to taking any risk and
keeps his entire wealth in cash (or invested at the riskless return). Other investors will
recognize that taking some risks by investing in the risky asset might provide 
gain opportunities above the riskless return. But this has a cost, the risk, and the question
is how to choose the optimal degree of risk/return.

Let us evaluate the value $S(t)$ of the portfolio comprising a fraction $f$ of the risky asset
after $t$ time steps. If we note $r_1, r_2,..., r_t$ the specific returns observed during these
$t$ time steps, the wealth becomes
\be
S(t) = \prod_{j=1}^t \biggl(1-f+fe^{r_j}\biggl) = \prod_{j=1}^t \biggl(1+(e^{r_j}-1)f\biggl) .
\label{gaingen}
\ee
This expression contains all the information, as long as we can enumerate all possible
scenarios $r_1, r_2,..., r_t$. In the absence of correlation between the returns $r_j$, all the
information is in the distribution $P(S(t))$ of the wealth. The problem thus amounts to
determine $P(S(t))$ knowing the distribution $P(r)$ of the one-step-returns of the risky
asset. 

Consider first the situation where the returns
$r_j$ are small ($r_j<<1$ for all $j$), $e^{r_j}-1 \simeq r_j$
and $1+r_j f \simeq e^{r_j f}$. The expression (\ref{gaingen}) becomes
\be
S(t) \simeq \prod_{j=1}^t  e^{r_j f} = \exp\biggl( f \sum_{j=1}^t r_j \biggl) .
\ee
${1 \over f} \log S(t)$ is the sum of $t$ random variables and the central limit theorem gives
\be
P(S(t)) \simeq {1 \over S_t \sqrt{2 \pi t \sigma^2}} \exp \biggl(-{(\log S_t - mft)^2 \over
2tf^2 \sigma^2} \biggl) ,
\label{resultlognff}
\ee 
where $m \equiv \langle r_j \rangle$ and $\sigma^2 \equiv \langle r_j^2 \rangle - m^2$.
The mean return of this portfolio is $r(f)=mf$ and its variance per unit time is 
$v(f)=\sigma^2 f^2$. 

In this Gaussian limit, the risk is completely captured by the single number $v(f)$.
For a given return $r=mf$, we get  
\be
r = {m \over \sigma} \sqrt{v}~,
\label{erzts}
\ee
by eliminating $f$. This relationship (\ref{erzts}) sums up the standard result of the 
celebrated Markovitz portfolio theory \cite{Markowitz,Markowitzb}, 
namely the more risk one takes (quantified by 
$\sqrt{v}$, the larger the expected return $r$. There is no best optimal strategy
$f$ but rather a continuum of possibilities offered to the investor and whose choice
depends on his risk aversion.

In the expression (\ref{resultlognff}), $S(t)$ behaves typically
as $S(t) = e^{rt + \eta \sqrt{vt}}$, where $\eta$ is a gaussian random variable with 
zero mean and unit variance. $S(t)$ becomes larger than one with almost certainty for
times larger than the characteristic time
\be 
t^* = {2v \over r^2} ,
\label{tempscharafgf}
\ee
at which the mean return $rt$ becomes comparable with the amplitude $\sqrt{2vt}$ 
of the fluctuations. Thus, the decision to invest a fraction of his wealth in the risky asset
must take into account the time interval during which he decides to immobilize his investment
\cite{Winger}. If this interval is too short, the results will be dominated by the fluctuations
and the risk is large. The investment in the risky asset becomes profitable only for time
intervals larger than $t^*$. The existence of this temporal factor in the investment strategy 
can be embodied in the so-called Sharpe ratio defined by the ratio of the expected return over the
standard deviation
\be
Sharpe_1 = {r \over \sqrt{v}}~.
\label{sharpedefffgqf}
\ee
$Sharpe_1$ can also be expressed as $Sharpe_1 = \sqrt{2} [t^*(f)]^{-1/2}$. The Sharpe ratio
depends on the time scale. A good investment strives to decrease the time scale
$t^*(f)$ beyond which the return is not anymore sensitive to fluctuations, hence to increase
the Sharpe ratio.

\subsection{An illustrative case}

Before addressing the question of large risks, let us make concrete the previous results
by specifying the distribution $P(r_j)$ of the risky asset. We consider a binomial case
which has the advantage of allowing for an exact treatment. The return of the risky asset
is thus assumed to be either
\be 
e^{r_j} = \lambda > 0  \qquad \qquad {\rm \ with \ the \ probability } ~~~ p ,
\label{rtigjjdfghf}
\ee
or
\be
e^{r_j} = 0  \qquad \qquad {\rm \ with \ the \ probability }~~~ 1-p .
\label{modelsimpledqd}
\ee
This particular case can be interpreted as a casino game and has first been studied by Kelly
 \cite{Kelly}, in relation with the theory of information which allows one to derive 
naturally the mean-variance approach in the Gaussian world.

A unit currency, from which a fraction $f$ is invested in the risky asset, becomes on average
after one time step\,: 
\be 
p [1+(\lambda - 1)f] + (1-p)[1-f]  = 1 +  (p\lambda - 1)f ~.
\label{gaindollar}
\ee
It is clear that only the case $p\lambda - 1 > 0$ is interesting and lead to investment 
strategies with  $f \neq 0$. The average return per time step is
$\log [1 +  (p\lambda - 1)f]$ while the average gain is $(p\lambda - 1)f$.

In this model, the scenario $r_1, r_2,..., r_t$ reduces to the knowledge of the number
$j$ of times where a gain (\ref{rtigjjdfghf}) was realized. The case
(\ref{modelsimpledqd}) for a loss then occurred $t-j$ times. The expression
(\ref{gaingen}) becomes
\be
S_j(t) = \biggl(1 + (\lambda - 1)f \biggl)^j (1-f)^{t-j} ~ .
\label{valeur}
\ee
The probability that this wealth is realized is simply the binomial law
\be
{\cal P}(j) = {t \choose j} p^j (1-p)^{t-j} ~.
\label{probaa}
\ee
The problem of the optimizing the investment is completely captured by the two formulas
(\ref{valeur}) and (\ref{probaa}). In the limit of large investment times, the Stirling
formula leads to the following expression
\be
{\cal P}(j) \simeq {1 \over \sqrt{2\pi tp(1-p)}} e^{-{(j-pt)^2 \over 2tp(1-p)}} ~, 
\label{probab}
\ee
valid close to its maximum.
This expression (\ref{probab}) is nothing but the translation of the central limit theorem
applied to the Binomial law, where the sum is of the type $\sum_{i=1}^t x_i$ with
$x_i=1$ with probability $p$ and $x_i=0$  probability $1-p$.
Expressing $j$ as a function of $S_j(t)$ in
(\ref{valeur}),  
\be 
j={\log{S_j(t) \over (1-f)^t} \over \log{1 + (\lambda - 1)f \over 1-f}}~,
\ee
we obtain the (log-normal) distribution of portfolio wealths at time $t$
\be
{\cal P}(S(t))  \simeq {1 \over \sqrt{2\pi tp(1-p)}} \exp \biggl(-{(\log {S(t) \over
S_{pp}(t)})^2 \over 2 t v(f)} \biggl) ~,
\label{probac}
\ee
where
\be
S_{pp}(t) = \biggl(1 + (\lambda - 1)f \biggl)^{pt} (1-f)^{(1-p)t}  
\label{plusproba}
\ee
is the most probable value of $S(t)$, i.e. the wealth that maximize the probability.
It simply corresponds to the most probable value of $j$ equal to $pt$. The variance 
$v(f)$ is given by
\be
v(f) = p(1-p) \biggl(\log{1 + (\lambda - 1)f \over 1-f}\biggl)^2 ~ .
\label{variance}
\ee
In this Gaussian limit, the portfolio selection problem is completely embodied by the 
two variables\,: the average return $r(f)$ defined by $r(f)={\log S_{pp}(t) \over t}$,
i.e. $S_{pp}(t)=e^{rt}$, giving 
\be
r(f) = p \log (1 + (\lambda - 1)f) + (1-p) \log (1-f) , 
\label{rendement}
\ee
and the variance $v(f)$. Eliminating $f$ allows us to obtain the mean-variance representation 
$r(v)$ which has become standard in portfolio theory.

Figure 1 shows $r$ and $v$ as a function of $f$ for $p=1/2$ and $\lambda = 2.1$. These
numerical values give an average gain (\ref{gaindollar}) $p\lambda - 1$ of $5 \%$.
Figure 2 gives $r$ as a function of $v$ for the same values. For $f$ small, 
$r \simeq (p\lambda - 1) f$ and $v \simeq p(1-p) \lambda^2 f^2$, which retrieves the
dependence $r \propto \sqrt{v}$ found in the preceeding section. 
Figure 3 shows as a function of $f$ for $r \geq 0$ the behavior of $t^*(f)$ defined by
(\ref{tempscharafgf}) and of $Sharpe_1 = \sqrt{2} [t^*(f)]^{-1/2} = {r \over \sqrt{v}}$
defined for a unit time step. It is profitable to invest in the risky asset only if one
is ready to freeze the investment over a period larger than $t^*(f)$. This time dimension is
often forgotten or dismissed in the more standard approach that does not distinguish between the
most probable and the mean gain. If the investment period $t$ is fixed, the best portfolios 
correspond to those with $t^*(f) < t$. This introduces the constraint $r > \sqrt{v \over t}$. 

We observe that, for $p\lambda > 1$, $r$ exhibits a maximum
\be
r_{max} = p \log \lambda p + (1-p) \log {\lambda (1-p) \over \lambda - 1}
\label{gain}
\ee
reached for an optimal fraction $f^*$ invested in the risky asset which is equal to
\be
f^*={p\lambda - 1 \over \lambda - 1} .
\label{azdfgw} 
\ee
For $p=1/2$ and $\lambda = 2.1$, $f^* = 4.54 \%$, $r_{max} = 0.113 \%$ and $v(f^*) =
0.00227$. The volatility $\sqrt{v}$, which gives the order of magnitude of the fluctuations of
$r$, is equal to $4.76 \%$, i.e. $t^*(f^*) \simeq 3500 \tau$ and $Sharpe_1(f^*) =
0.024$. For $p=1/2$ and $\lambda = 2.5$, $t^*(f^*)$ becomes
smaller than $200 \tau$, corresponding to $Sharpe_1 = 0.07$, with an optimum return better than
$2 \%$. For $p=0.9$ and $\lambda = 1.2$,  $t^*(f^*)$ is also of the order of
$200 \tau$, corresponding to $Sharpe_1 = 0.07$, with a similar optimum return 
$2 \%$. As $p\lambda$ becomes much larger than $1$, $f^*$ increases as it becomes
more and more interesting to invest in the risky asset. For $p=1/2$ and $\lambda = 5$, 
$r_{max} = 2.5 \%$, $f^* \simeq 0.4$,  $t^*(f^*) \simeq 20$ and $Sharpe_1 = 0.25$.
For $\lambda \rightarrow \infty$ and in the regime of positive mean returns, 
$Sharpe_1$ becomes independent of $f$ and of $\lambda$: $Sharpe_1 \rightarrow \sqrt{p \over
1-p}$. It becomes completely controlled by the stochastic part of the process.

The existence of a maximum (\ref{gain}) was not predicted by the linear analysis of the previous
section. It stems, as we now analyze, from the existence of large deviations inherent in the
multiplicative processes (\ref{gaingen},\ref{valeur}).

\subsection{Large risks and large deviations}

\subsubsection{Typical versus average returns}

It is a well-known fact that average and typical values can 
be drastically different in multiplicative processes. This difference is at the 
origin of the existence of a maximum (\ref{gain}).

Let us come back to the exact expressions (\ref{valeur}) and (\ref{probaa}). The 
mathematical expectation $\langle S(t) \rangle$ of the value $S(t)$ of the portfolio
corresponds to taking the average
over a large number of realizations, in other words over a large number of investors. This
will be a useful quantity for a bank, say, as opposed to a single investor.
It is given by
\be
\langle S(t) \rangle = \sum_{j=0}^t  {\cal P}(j) S_j(t)  .
\ee
This expression can be summed and yields
\be
\langle S(t) \rangle = \biggl( p [1+(\lambda - 1)f] + (1-p)[1-f] \biggl)^t = 
\biggl( (p\lambda - 1)f + 1 \biggl)^t  .
\label{som}
\ee
We thus simply retrieve the expression (\ref{gaindollar}) taken to the power
$t$, i.e. 
\be
(1 + \langle {\rm gain ~over ~one ~time ~step}~ \rangle)^t~.
\ee
$\langle S(t) \rangle$ is not the same as the most probable value
$S_{pp}(t)$ given by (\ref{plusproba}). 
$\langle S(t) \rangle$ and $S_{pp}(t)$ have very different behaviors.
Figure 4 gives the dependence of 
$\langle S(1) \rangle$ and $S_{pp}(1)$ as a function of $f$ for the case
 $p=1/2$ and $\lambda = 2.1$ studied previously. As can be seen from
(\ref{som}), $\langle S(1) \rangle$ is a linear function of $f$ while
$S_{pp}(1)$ exhibits a highly nonlinear dependence. Technically, the difference
can be tracked back to the fact that $S_{pp}(t)$ 
is the exponential of the average of the logarithm of $S(t)$, i.e. the geometrical mean of
$S(t)$\,:  
\be 
S_{pp}(t) = e^{\langle \log S(t) \rangle}  ,
\label{logmoy}
\ee
which can checked by calculating  $\langle \log S(t) \rangle = \sum_{j=0}^t  {\cal
P}(j) \log S_j(t)$. This is to be compared to
\be
\langle S(t) \rangle = \langle e^{\log S(t)} \rangle ,
\label{moyexp}
\ee
which is the arithmetic average. For $f$ small for which the return is close to zero, the 
two values become indistinguishable, as can be seen in figure 4. $f<<1$ implies small 
fluctuations of wealth and the two averages become identical in this limit.

In contrast for arbitrary $f$, $\langle \log S(t) \rangle$ and
$\log \langle  S(t) \rangle$ can be very different.
 Mathematically, the exchange between the mean and the exponentiation is not justified.
Expanding the exponential in (\ref{moyexp}) yields
\be
\langle S(t) \rangle = 1 + \langle \log S(t) \rangle + {1 \over 2} 
\langle (\log S(t))^2 \rangle +... ~,
\ee
that can be compared with
\be
S_{pp}(t) = e^{\langle \log S(t) \rangle} = 1 + \langle \log S(t) \rangle + {1 \over 2} 
\langle \log S(t) \rangle^2 +...  ~.
\ee
The difference between the two quantities occurs at the second order in the expansion.
More generally, for any positive random variable, $\langle x^k \rangle > \langle x \rangle^k$ and
thus  $\langle S(t) \rangle > S_{pp}(t)$. 
As an illustration, figure 4 shows that 
$\langle S(1) \rangle \simeq 1.002$ while $S_{pp}(1) \simeq 1.001$ at the maximum of the mean return.
After $1000$ time steps, this leads to
$\langle S(1000) \rangle \simeq 7.4$ while $S_{pp}(1000) \simeq 2.7$. The difference explodes 
exponentially at large times. Notice that, for $f > 0.09$, 
$r$ becomes negative and thus $S_{pp}(t)$ decreases exponentially while
$\langle S(t) \rangle$ still increases exponentially\,: a typical investor will loose almost
surely while very rare investors will exhibit shameless gains of such a magnitude as to make
the average economy still growing!

The origin of these differences can be found in the large fluctuations of multiplicative
process, exhibiting rare fluctuations but of sufficient magnitude to control the behavior of the 
mean gain. For instance, a sequence of positive returns such that there are $j=t$ successes
occurs with an exponentially small probability. However, its contribution to the mean gain
has an exponentially large magnitude. As a consequence, this type of scenarios brings in
an important contribution to $\langle S(t) \rangle$ and a dominant contribution of higher moments.

The difference between (\ref{logmoy}) and (\ref{moyexp}) illustrates dramatically the
importance of diversification. The result (\ref{moyexp}) quantified the results obtained
by an investor $A$ who could divide his initial wealth in $N$ independent assets, each asset
corresponding to the couple ($f$ invested in the risky game, $1-f$ kept at the riskless return),
each playing a given game independently of the others. This diversification is gratifying as
his total wealth is controlled by the mean gain, this becoming true in the limit of large $N$. This 
is very different from the result (\ref{logmoy}) 
obtained by an investor $B$ who has only a single asset
and is thus completely controlled by a single scenario. 
In practice, for the wealth to be described by (\ref{moyexp}), 
the number of assets $N$ must be exponentially large in $t$ \cite{Redner}\,:
$N \propto e^{ct}$, where $c>0$ is an increasing function of $f$. At short times, it is not
very useful to diversify, but as time passes by, the diversification becomes essential. 
This introduces an additional level of complexity\,: for practical implementation, $N$ can not
be arbitrarily large\,; one can thus imagine investment strategies which modify $f$ such that 
the number $N(t)$ necessary to track the mean gain remains bounded.

These effects are well-known in the physics of disordered media and are at the origin of 
a wealth of effects, culminating in the non-ergodic behavior of spin-glasses. The consequence
of large fluctuations in multiplicative processes in the context of portfolio selection has
also been stressed recently \cite{Zhangmaslov}.

\subsubsection{Large deviations}

Large risks are characterized by the
existence of deviations from the behavior described by the central limit theorem.

Indeed, the previous considerations have relied on the fact that the fluctuations of the 
return could be adequately quantified by a unique number, namely the variance. 
For those fluctuations of $S(t)$ away from the most probable value
$S_{pp}(t)$ given by (\ref{plusproba}) that are larger than the standard deviation of the
Gaussian approximation (\ref{probac}), the average return and the variance are no more 
enough to quantify the whole spectrum of risks. This problem is very important because, 
as already pointed out, there is not a unique optimal investment strategy but rather a set
of strategies each corresponding to a given risk aversion. Since risk aversion relies on
a subtle combination of psychological and financial considerations, it is very important
to provide a reliable appreciation of all the possible dimensions of the risks.

Technically, an answer is provided by the theory of large deviations, which establishes the
probability, for large $t$, for the mean of independent and identically distributed
random variables\,:
\be 
Prob[ {1 \over t} \sum_{j=1}^t m_j \simeq x ] \sim e^{ts(x)} ~ ,
\label{cramera}
\ee
where $s(x)$ is the Cram\'er function \cite{Lanford,Frisch,nousfrisch}.
When applied to the binomial law (\ref{probaa})
where the sum is of the type $\sum_{i=1}^t x_i$ with $x_i=1$ with probability $p$ and $x_i=0$
with probability $1-p$,  this result is nothing but the one obtained from the Stirling approximation
of ${t \choose j}$\,:
\be
{\cal P}(j=xt) = e^{ts(x)}  ,
\label{cramerb}
\ee
with
\be
s(x) = x \log p + (1-x) \log (1-p) -x \log x - (1-x) \log (1-x) ~, 
\qquad {\rm \ for \ 0 < x < 1}  
\label{cramerc}
\ee
\be
s(x) = - \infty  \qquad \qquad {\rm \ otherwise~.}
\ee
Notice that $s(x)$ reduces to $s(x) = -{(x-p)^2 \over 2p(1-p)}$ in the neighborhood of
its maximum $x=p$, which retrieved exactly the expression (\ref{probab}) used until now.
Figure 5 shows $s(x)$ and its parabolic approximation for $p=0.5$ and $p=0.95$.
In the first case $p=0.5$, the parabolic approximation, leading to the Gaussian distribution
and to all the previously quoted results, overestimates the risks. For instance, the 
risk to loose systematically at each time step corresponds to $x=0$, i.e.
${\cal P}(j=0) = e^{ts(0)}$, where $s(0) \simeq -0.5$ according to the parabolic approximation,
which can be compared to the exact value $-0.7$. The probability to loose at each time step
during $t$ is thus estimated as $e^{-0.5t} \approx 7~10^{-3}$ and $e^{-0.7t} \approx 10^{-3}$ 
respectively for $t=10$, leading to an overestimation of this risk by a factor $7$ in the Gaussian 
approximation. For $p>0.7$, the effect goes in the other direction. For instance, for $p=0.95$, we 
see in the figure 5b that the function $s(x)$ is significantly above its parabolic approximation.
As a consequence, large deviations are much more probable than predicted by the Gaussian 
approximation and the standard mean-variance theory.

In order to quantify this result, we express (\ref{valeur}) as
\be
S_x(t) = \biggl( (1-f)^{1-x} [1 + (\lambda - 1)f]^x \biggl)^t  ~,
\label{rendex}
\ee
using $j=xt$. The distribution ${\cal P}(x)$ is given by (\ref{cramerb}) with
(\ref{cramerc}). ${\cal P}(x)$ with the expression (\ref{rendex}) provides the distribution of
$S(t)$ by eliminating $x$. We can also use these two formulas by using $x$ as a parameter\,:
for $S(t)$ given, we determine the corresponding $x$ from 
(\ref{rendex}) which is then reported into (\ref{cramerb}) with
(\ref{cramerc}) to derive the probability. Recall that $x$ is simply the frequency of gains.
Large deviations correspond to a frequency $x$ different from $p$ by an amount of order one.

As an illustration, let us take $p=0.95$ and $\lambda=1.1$, for whose realistic behaviors are
found ($5 \%$ probability to loose and a gain of $10 \%$). Let us assume that the investor chooses 
$f=0.4$ (which is close to the maximum) which corresponds to an average return
$r(f=0.4) = 1.2 \%$. This result is also obtained from (\ref{rendex}) with $x=p=0.95$, 
using the definition $S(t)= e^{rt}$. What is the probability that, instead of getting the
expected return, this investment looses on average $-1.6 \%$?
This value corresponds to $x=0.9$ as seen from (\ref{rendex}).
The calculation of $s(x=0.9)$ gives $-0.0264$ in the parabolic approximation and
$-0.0206$ with the exact expression (\ref{cramerc}). The probability of this bad luck is thus
$e^{-0.0264t}$ in the Gaussian approximation and $e^{-0.0206t}$ according to the 
Cram\'er expression. For $t=100 \tau$, we find respectively a probability equal to $7 \%$ 
and $13 \%$. Thus, the Gaussian approximation under-estimates by almost a factor of two
the probability of such a scenario.

Consider another example\,: what is the probability that the average return be equal to
$-0.5 \%$? Following the previous steps, we find that this return corresponds to
$x \simeq 0.92$ for which $s(x=0.92)=-9.5 \cdot 10^{-3}$ in the parabolic approximation and
$s(x=0.92)=-8.1 \cdot 10^{-3}$ for the exact expression (\ref{cramerc}).
For $t=1000 \tau$, we obtain $7.5 \cdot 10^{-5}$ and $3 \cdot 10^{-3}$ respectively for the 
probability to get such a return. Here, the Gaussian approximation is off by a factor $40$!

To sum up the section, it is not possible in general to capture the degree of risk to which
one is exposed to by the single variance parameter. A natural extension is to consider the full
Cram\'er function as the quantity encoding the risk. The choice of the best investment becomes
then even more dependent on additional considerations involving the aversion to rare and 
relatively large risks. 

Is it possible to choose and investment that minimize the large risks? In the present simple model,
the answer is\,: choose $f = 0$ is you are afraid of large risks. To see it, we pose the problem
in the following way\,: we try to minimize the probability that the return be less than some
level $\rho$, whose specific value is investor-dependent and not specified as long as it is 
sufficiently in the tail. This level $\rho$ of a minimum tolerable return corresponds to a 
frequency $x_{\rho}(f)$ by the equation
\be
(1-f)^{1-x_{\rho}} [1 + (\lambda - 1)f]^{x_{\rho}} = e^{\rho} ~.
\label{gghjfkdq}
\ee
With (\ref{variance}), it gives $x_{\rho}$ as
\be
x_{\rho} = \sqrt{p(1-p)} \biggl( {\rho - \log (1-f) \over \sigma (f)} \biggl) ,
\label{xxxxxccvf}
\ee
where $\sigma (f) = \sqrt{v(f)}$ is the standard deviation. The worst return is of course
$\log (1-f)$ for which $x$ is zero. Notice that 
the value $x_{\rho} = p$ corresponds to the typical return $r(f)$ given by
(\ref{rendement}). For $\rho < r(f)$, the frequency of gains $x_{\rho}$ is less than $p$.

The probability to observe a return less or equal to $\rho$ is $\int_0^{x_{\rho}}
P(x) dx \equiv \sum_{j=0}^{t x_{\rho}} {t \choose j} p^j (1-p)^{t-j}$, in which each term
behaves like $e^{ts(x=j/t)}$. Is there an optimal $f$ such that 
$\int_0^{x_{\rho}} e^{ts(x_{\rho})}$ is minimum?
Using (\ref{cramerc}), the extremum of $\int_0^{x_{\rho}} e^{ts(x)} dx$ with respect to
$f$ is given by the condition $0 = {d x_{\rho} \over d f} = -{1-x_{\rho} \over 1-f} + {(\lambda -
1) x_{\rho} \over 1 + (\lambda - 1) f}$. However, this corresponds to a maximum
since the second derivative is negative. The value $f_1$ which makes vanish ${d x_{\rho} \over d f}$
corresponds to a maximum of probability to gain a severa loss. These values are thus to be 
avoided. With  (\ref{xxxxxccvf}), they are $1-f = {\lambda \over \lambda - 1} (1-x_{\rho})$
with $\rho = \log {\lambda \over \lambda - 1}  + (1-x_{\rho}) \log(1-x_{\rho}) +
x_{\rho} \log x_{\rho} + x_{\rho} \log (\lambda - 1)$.
$(1-x_{\rho}) \log(1-x_{\rho}) + x_{\rho} \log x_{\rho}$ being negative with a minimum
$-\log 2$ for  $x_{\rho} = 1/2$ and vanishing for $x_{\rho} = 0$ and $1$, there are
non-trivial solutions for $x_{\rho} < p$. There exists other situations where an optimal 
investment can be devised such as to minimize the probability of large losses, as we discuss below.

\section{Portfolios made of uncorrelated assets}

Consider the sum
\be 
S(t) = \sum_{i=1}^N p_i x_i(t) ~,
\label{sumport}
\ee
where $x_i(t)$ represents the value at time $t$ of the $i$-th asset
and $p_i$ is its weight in the portfolio. By normalization, we have $\sum_{i=1}^N p_i = 1$. 
$S(t)$ is the total value of the portfolio made of $N$ assets at time $t$.

\subsection{Gaussian limit and measure of risk by the variance}

We would like to characterize the probability distribution $P_S(S(t))$ of $S(t)$, knowing the
distributions $P_i(x_i)$ of the $x_i$ for the different assets. For uncorrelated assets, the
general formal solution reads
\be
P_S(S) = \int dx_1 P_1(x_1) \int dx_2 P_2(x_2) ... \int dx_j P_j(x_j) ...
\int dx_N P_N(x_N)  \delta\biggl(S(t) - \sum_{i=1}^N p_i x_i(t)\biggl) ~ .
\label{solut}
\ee
Taking the Fourier transform 
$\hat P_S(k) \equiv \int_{-\infty}^{+\infty} dS P_S(S) e^{-ik S}$ of 
(\ref{solut}) gives, by the definition of the characteristic function \cite{Pury}, 
\be
\hat P_S(k) = \hat P_1(p_1k) \hat P_2(p_2k)...\hat P_j(p_j k)...
\hat P_N(p_N k)  ~.
\label{fourier}
\ee
This equation expresses that the Fourier transform of $P_S(S)$ is the product of the Fourier
transform of the distribution of each constituent asset with an argument proportional to 
their respective weight in the portfolio. We now use the cumulant expansion of the characteristic
function\,:
\be
\log \hat P_j(p_j k) = -ik p_j C_1(x_j) - {1 \over 2} p_j^2 k^2 C_2(x_j) + 
{\cal O}(p^3 k^3) ~,
\label{cumul}
\ee
where $C_1(x_j) \equiv \langle x_j \rangle$ is the average of $x_j$ and $C_2(x_j) \equiv 
\langle x_j^2 \rangle -  \langle x_j \rangle^2$ is the variance of $x_j$. We assume
for the time being that these cumulants exist and are finite. The notation
${\cal O}(p^3 k^3)$ means that the higher order terms are at least of order 
$p^3 k^3$. In the limit where the number $N$ of assets is very large, the weights 
$p_i$ are of order ${1 \over N}$ (we avoid the special cases where one or a few assets
are predominant). In this case, the previous cumulant expansion is warranted since all
arguments are small and, keeping terms up to the quadratic order, we obtain
\be
\hat P_S(k) \simeq \exp[-ik \sum_{j=1}^N p_i C_1(x_j) - 
{1 \over 2} k^2  \sum_{j=1}^N  p_j^2 C_2(x_j) ]~.
\label{produ}
\ee
Its Fourier inverse is
\be
P_S(S) \simeq {1 \over \sqrt{2 \pi V}} \exp[-{(S-\langle S \rangle)^2 \over 2V}] ~ ,
\label{gauss}
\ee
where the average of $S$ is
\be 
\langle S \rangle = \sum_{j=1}^N p_j \langle x_j \rangle  ~,
\label{moyenne}
\ee
et la variance $V$ de $S$ s'\'ecrit
\be
V = \sum_{j=1}^N  p_j^2 C_2(x_j) ~ .
\label{varianc}
\ee
These expressions express again the validity of the central limit theorem.

\subsection{Large risks\,: the case of exponentially distributed assets}

The previous Gaussian approximation (\ref{gauss}) does not apply for large risks.
We have already met this fact in the binomial model with the difference between 
(\ref{probac}) and (\ref{cramerb}, \ref{cramerc}). We now present a slightly more
general illustration of this fact by studying the case of assets that are 
exponentially distributed. It turns out that this situation is not far from 
reality when one deals with daily returns \cite{Laherrere}.

\subsubsection{Determination of the distribution of portfolio values}

To simplify, we assume that the distributions of asset return is symmetric\,:
\be 
P_j(\delta x_j) = {1 \over 2} \alpha_j e^{\alpha_j |\delta x_j|} ~ ,
\label{expon}
\ee
for $-\infty < \delta x_j < + \infty$, where $\langle \delta x_j \rangle = 0$
and the variance $\delta x_j$ is $C_2(v_j) = {1 \over \alpha_j^2}$.

The Fourier transform of (\ref{expon}) is
\be 
\hat P_j(k) = {1 \over 1 + (k\alpha_j^{-1})^2}~.
\label{lorenz}
\ee
Inserting this expression in (\ref{fourier}) and taking the inverse Fourier transform, we get
\be
P_S(\delta S) =  {1 \over 2\pi} \int_{-\infty}^{+\infty} dk {e^{ik \delta S} \over 
\prod_{j=1}^N [1+(kp_j\alpha_j^{-1})^2]}  ~.
\label{produit}
\ee
We retrieve the Gaussian approximation by noting that
\be
{1 \over 1+(kp_j\alpha_j^{-1})^2} = \exp \biggl(-\log [1+(kp_j\alpha_j^{-1})^2]\biggl)
\simeq \exp \biggl(- (kp_j\alpha_j^{-1})^2\biggl) ,
\label{devel}
\ee
and thus
\be 
P_S(\delta S) \simeq {1 \over 2\pi} \int_{-\infty}^{+\infty} dk e^{ik \delta S} 
\exp \biggl(- k^2 \sum_{j=1}^N (p_j\alpha_j^{-1})^2\biggl)~,
\ee
which recovers (\ref{gauss}).

The integral (\ref{produit}) can be performed exactly by using Cauchy's theorem\,: one replaces
the integral over the interval $-\infty < k < + \infty$ by an integral over a contour 
in the complex $k$ plane. This contour is formed of the real axis and closes itself by a half-circle
of infinite radius in the top half-plane. Cauchy's residue theorem then gives
\be 
P_S(\delta S) = {1 \over 4} \sum_{j=1}^N {1 \over p_j\alpha_j^{-1}}
{1 \over \prod_{i \neq j} \biggl(({p_j\alpha_j^{-1} \over p_i\alpha_i^{-1}})^2 - 1 \biggl)}
\exp [-{ |\delta S| \over p_j\alpha_j^{-1}}]  ~.
\label{resul}
\ee
This result is correct only if all the values $p_i \alpha_i$ are different. The special 
cases where several values are the same do not pose particular problems and can also be
explicitely treated. This expression (\ref{resul}) shows that the large risks are given 
by 
\be  
P_S(\delta S)_{\delta S \rightarrow -\infty} \simeq {\hat \alpha \over 4
\prod_{i \neq j_{max}} \biggl(({p_{j_{max}}\alpha_{j_{max}}^{-1}  \over p_i\alpha_i^{-1}})^2
- 1 \biggl)}  e^{-\hat \alpha |\delta S|} ~ ,  
\label{asymp}
\ee
where 
\be
\hat \alpha = {\alpha_{j_{max}} \over p_{j_{max}}}
\ee
with $j_{max}$ being the value of  $j$ which corresponds to the largest  $p_j\alpha_j^{-1}$. 
The order of magnitude of the largest fluctuations of $\delta S$ 
is thus $\hat \alpha^{-1}$, i.e. $\hat \alpha^{-1}$ 
provides a good estimation of the extreme risk of the portfolio.

\subsubsection{Diversification in the presence of extreme risks}

To protect oneself against the large risks, one should thus minimize
$\hat \alpha^{-1}$, i.e. solve the optimization problem on the weights
$p_1,p_2,...,p_N$, which consists first, for fixed weight $p_1,p_2,...,p_N$, in finding the 
smallest ratio ${\alpha_j \over p_j}$ that we note $Min_{j=1,...N} {\alpha_j \over p_j}$,
and then to determine the weights $p_1,p_2,...,p_N$ such that the smallest ratio be the largest
possible. To sum up, we thus search for the solution of the max-min problem
\be
Max_{p_1,p_2,...,p_N} Min_{(j=1,...N)} {\alpha_j \over p_j}~.
\ee
To solve this problem, we invoke the following identity
\be
Min_{(j=1,...N)} {\alpha_j \over p_j}
= lim_{q \to +\infty} \biggl( \sum_{j=1}^N ({\alpha_j \over p_j})^{-q} \biggl)^{-1/q}~ .
\ee
The idea is to invert the limit  $q \to +\infty$ and the maximization with respect to
the weights. At $q$ fixed, we then maximize this expression with respect to the weights
$p_j$ together with the normalization constraint $\sum_{j=1}^N p_j = 1$, by using
Lagrange multiplier method. The solution is $({\alpha_j \over p_j})^{-q} = $ constant,
independently of $j$. Then taking the limit $q \to +\infty$, the only possible solution is
$p_j \propto \alpha_j$, which gives 
\be
p_k = {\alpha_k \over \sum_{j=1}^N \alpha_j}~,
\label{resythgfglk}
\ee
after normalization. We will derive again this result below (see the equation (\ref{premstrat})) 
by a different argument in terms of the minimization of the portfolio kurtosis, thus providing
a different perspective to this result which constitutes the large deviation correction to 
the mean-variance approach for exponentially distributed assets. The same results have
been derived independently in Ref.\cite{Bouchaud}.

\subsubsection{Large risks and optimal portfolios}

A better measure of the risk than the sole knowledge of $\hat \alpha$ is provided by 
the probability that the loss $\delta S$ of the portfolio be larger than some threshold
$\lambda$. The parameter $\lambda$ is a priori arbitrary and is chosen by the investor in
view of his own risk aversion. $\lambda$ is a so-called VaR, or value-at-risk. It is the value
that can be lost at the probability level 
$\int_{-\infty}^{-\lambda} P_S(\delta S) dv$\,:
\be  
\int_{-\infty}^{-\lambda} P_S(\delta S) d\delta S = {1 \over 4}
\sum_{j=1}^N  {1 \over \prod_{i \neq j} \biggl(({p_j\alpha_j^{-1} \over p_i\alpha_i^{-1}})^2
- 1 \biggl)} \exp [-{ \lambda \over p_j\alpha_j^{-1}}]  .
\label{risq}
\ee
This expression provides the probability to observe a loss larger than $\lambda$ 
during the unit time step considered here. Reciprocally, we can choose the confidence interval,
say $95 \%$ and determine what is the threshold $\lambda$ such that the maximum loss remains
smaller than $\lambda$ in $95 \%$ out of all scenarios. The VaR $\lambda$ is solution of
\be
\int_{-\infty}^{-\lambda} P_S(\delta S) d\delta S = 0.05~,
\ee
which can be explicitely solved using (\ref{risq}).

Let us now determine the weights of the portfolio that minimize the probability of loss
larger than $\lambda$. The expression (\ref{risq}) is not very convenient for an 
analytical calculation. We thus turn to a more robust measure of the VaR given by 
the fact that a smooth estimation of 
$\int_{-\infty}^{-\lambda} P_S(\delta S) d\delta S$ is provided by $\int_{-\infty}^0
P_S(\delta S) (1 - e^{\delta S \over \lambda}) d\delta S$. Indeed, $1 - e^{\delta S \over
\lambda}$ is close to $1$ for  $\delta S < - \lambda$ and is negligible in the other case.
$(1 - e^{\delta S \over \lambda})$ plays a role similar to a utility function, quantifying the
sensitivity of the investor to large fluctuations. This leads to the following smooth 
definition of the Var $\lambda$\,:
\be 
TL_{\beta=0}\biggl(P_S(|\delta S|)\biggl) - TL_{\beta={1 \over
\lambda}}\biggl(P_S(|\delta S|)\biggl) = 1-p ,
\label{laplcnnd}
\ee
where $TL_{\beta } \biggl(f(x)\biggl) \equiv \int_0^{\infty} e^{-\beta x} f(x) dx$ is the 
Laplace transform of the function $f(x)$. The first term
$TL_{\beta=0}\biggl(P_S(|\delta S|)\biggl)$  is nothing but the total probability for 
a loss to occur (irrespective of its amplitude). This expression 
(\ref{laplcnnd}) has two advantages\,: i) a more progressive interpolation of the losses
to determine the VaR and ii) the use of the Laplace transform which can be directly
estimated in the case of portfolios from the product of individual Laplace transforms of the 
distribution of each asset.

Let us change variable and write $p_j \equiv \rho_j^2$, where the $\rho_j$ are the novel
parameters over which to minimize. This change of variables ensures that the weights remain
positive. Notice that we could relax this constraint and allow for negative weights which would
correspond to so-called ``short'' positions. We thus would like to minimize
\be
TL_{\beta=0}\biggl(P_S(|\delta S|)\biggl) - TL_{\beta={1 \over
\lambda}}\biggl(P_S(|\delta S|)\biggl) - \gamma \sum_{j=1}^N \rho_j^2
\label{lafgaeplcnnd}
\ee
with respect to the $\rho_1, \rho_2, ...., \rho_N$. $\gamma$ is a Lagrange parameter
that ensures the normalization of the weights. Using the analytic expression of the 
Laplace transforms, we can write (\ref{lafgaeplcnnd}) as
\be
1 - p(\lambda) - \gamma \sum_{j=1}^N \rho_j^2 ,
\label{laawplcnnd}
\ee
where 
\be
p(\lambda) \equiv {\prod_{j=1}^N \alpha_j  
\lambda \over \prod_{j=1}^N (\alpha_j \lambda + \rho_j^2)}~.
\ee
Putting to zero the derivative of (\ref{laawplcnnd}) with respect to each
$\rho_k$ gives
\be
p_k = {1 \over N} + {\lambda \over N}~ \sum_{j=1}^N (\alpha_j - \alpha_k)~.
\label{awqzp}
\ee
This solution exists provided that $\lambda$ is not too large such that the $p_k$'s 
remain positive. $\gamma$ is eliminated by the normalization.

For large $\lambda$,
\be
1 - p(\lambda) \approx \sum_{j=1}^N {p_j \over \alpha_j}~,
\ee
which recovers the situation treated in section 3.2.2.

\subsection{Beyond the Gaussian limit\,: cumulant expansion and large deviation theory}

For arbitrary assets with finite variance, the VaR at the probability level $p$ is given
by 
\be
\int_{-\infty}^{-\lambda} P_S(\delta S) d\delta S = 1-p ~.
\ee
The large deviation theorem allows us to get 
\be
P_S(\delta S) \sim e^{N s(\delta S)} ~ ,
\label{granddev}
\ee
where the Cram\'er function $s(x)$ can be expressed in terms of all the distributions
$P_j(\delta x_j)$ for $j=1$ to $N$ \cite{Lanford}
\be
s(\delta S) = Inf_{\beta} \biggl({1\over N} \sum_{j=1}^N \log \hat P_j(p_j\beta) + 
\beta \delta S \biggl) ~.
\label{premieb}
\ee
$Inf_{\beta}$ expresses the fact that we evaluate the term within the parenthesis for the 
value of $\beta$ which minimizes it. This expression (\ref{premieb}) together with (\ref{granddev}) 
gives the distribution of deviations that can take arbitrarily large values, i.e. much beyond
the Gaussian approximation. 

This expression also contains the Gaussian limit valid for 
small fluctuations. This can be seen from the formula (\ref{cumul}) adapted to the Laplace
transform (by replacing $ik$ by $\beta$). Noting $\langle \delta S \rangle = 
\sum_{j=1}^N p_j C_1(\delta x_j)$ and
$V = \sum_{j=1}^N p_j^2 C_2(\delta x_j)$ (equations (\ref{moyenne}, \ref{varianc})),
the expression  (\ref{premieb}) becomes
\be
N s(\delta S) = Inf_{\beta} (-\beta \langle \delta S \rangle + {1 \over 2}  V \beta^2 + 
\beta \delta S)~.
\ee
$s(\delta S)$ can then be obtained as the solution of a simple quadratic minimization. 
The value of $\beta$ that minimizes the expression within the parenthesis is 
\be
\beta = - {\delta S - \langle \delta S \rangle \over  V}~.
\ee
We thus finally obtain
\be
s(\delta S) = - {(\delta S - \langle \delta S \rangle)^2 \over 2 N V}~.
\ee
Reporting in (\ref{granddev}), we thus retrieve the Gaussian approximation.

Let us now use (\ref{premieb}) to express the first leading corrections to the Gaussian
approximation (\ref{gauss}). In this goal, we expand $\log \hat P_j(p_j\beta)$ 
on the cumulants $c_n^j$:
\be
\log \hat P_j(\beta) = \sum_{n=1}^{\infty} \frac{ c_n^j}{n!}(-\beta)^n ~.
\label{cumulantt}
\ee
Interchanging the sums over the assets $j$ and over the cumulants $n$, we write
(\ref{premieb}) as
\be
N s(\delta S) = Inf_{\beta} \biggl( \beta (S-\langle S \rangle) + {1 \over 2} V \beta^2 
- {1 \over 6} C_3 \beta^3 + {1 \over 24} C_4 \beta^4 + ... \biggl)  ~,
\label{cumulbb}
\ee
where $\langle S \rangle$ and $V$ are given by (\ref{moyenne}) and
(\ref{varianc}),
\be
C_3 = \sum_{j=1}^N  p_j^3 c_3^j ~ ,
\label{ccc}
\ee
and
\be
C_4 = \sum_{j=1}^N  p_j^4 c_4^j  ~.
\label{cccc}
\ee
If the distributions of the price variations of the assets are non-symmetric, then 
$C_3 \neq 0$ and the first correction to the Gaussian approximation reads
\be
P_S(S) \simeq \exp [-{(S-\langle S \rangle)^2 \over 2V}
\biggl(1 - {C_3 (S-\langle S \rangle) \over 3V^2}\biggl)] ~ . 
\label{gaussccc}
\ee
For symmetric distributions such that $c_3^j = 0$, the leading correction is proportional to the
kurtosis $\kappa = {C_4 \over V^2}$\,:
\be
P_S(S) \simeq \exp [-{(S-\langle S \rangle)^2 \over 2V}
\biggl(1 - {5 C_4 (S-\langle S \rangle)^2 \over 12 V^3}\biggl)]  . 
\label{gaussddd}
\ee
For a typical fluctuation $S - \langle S \rangle \sim \sqrt{V}$, the relative size of the 
correction is order ${5 C_4 \over 12 V^2} = {5 \kappa \over 12}$. Notice the negative sign
of the correction proportional to $C_4$ which means that large deviations are more probables
than extrapolated by the Gaussian approximation.

Let us illustrate these results for the exponential distributions. Consider the simple case
where all assets have the same distribution, i.e. $\alpha_j=\alpha$ for all $j$'s. Let us also
take all weights $p_j$ equal to $1/N$. The expression (\ref{premieb}) then yields
\be 
s(\delta S) = \log (\alpha \delta S) + 1 - \alpha \delta S  .
\label{special}
\ee
We recover directly this result by using the exact relation (\ref{produit}). Indeed, the family
of Gamma functions is close under convoluation
(\cite{Feller}, p. 47). Applying this result to the variable $\delta S
= {1 \over N} \sum_{j=1}^N \delta x_j$, this explains the additional factor $N$ in the
exponential and the prefactor. We thus have
\be
P_S(\delta S) = {N \alpha \over  \Gamma(N)}  (N \alpha \delta S)^{N-1}
e^{-N \alpha \delta S}  .
\label{exactt}
\ee
The expression (\ref{special}) reported in (\ref{granddev}) yields indeed
(\ref{exactt}). This result shows that the extreme tail of the distribution remains
essentially of the exponentialform $e^{-N\alpha \delta S}$, in agreement with our previous
results.

The figure 6 illustrates these results by showing the function 
$s(y \equiv \alpha \delta S) = \log (y) + 1 - y$, in comparison to its
parabolic approximation $s_g(y) = -{1 \over 2} (y - 1)^2$. 
It is clear that the parabolic approximation is correct only for 
small deviations around the mean value $y=1$. For
$|y| = 3$, $s_g(y)$ is already twice as large as $s(y)$, leading to a severe 
under-estimation of large fluctuations by the Gaussian approximation. Notice also that
for large $y$, the function $s(y)$ becomes essentially parallel to the straight line of 
slope $-1$, thus justifying the asymptotic shape (\ref{granddev}) of the distribution tail.

\section{General analysis of portfolios}

Let us consider $N$ assets $X^a$, $a=1,.., N$, each of which has an instantaneous value
$x^a_k$ at time $k$. We now present the analysis of portfolio optimization in the presence
of large fluctuations and for the general case of correlated assets. Before reaching this goal,
we first recall the standard mean-variance approach valid within the Gaussian approximation.

\subsection{Correlation matrix between assets}

Within the Gaussian approximation, correlations between assets can be fully characterized
by the determination of the correlation matrix
\be
V_{ab} \equiv \langle \delta x^a \delta x^b \rangle ~,
\label{defmatcor}
\ee
which also reads $V \equiv \langle X^a X^{aT} \rangle$ in matrix notation. 
$X^a$ is the column vector with row element $x^a$, for $a=1$ to $N$ and the exponent
$^T$ stands for the transpose operation. By construction, 
$V_{ab}$ is symmetric and for non-singular cases can be diagonalized with all its 
eigenvalues $\lambda_k$ being real. The elements of the eigen-vectors 
$P_{\alpha}$ are also real \cite{Healy}. Notice that $V_{ab}$ is a symmetric matrix with 
only $N$ and not $N(N+1)/2$ independent degrees of freedom. There is thus an orthogonal matrix 
$A$ such that $A^T V A = \lambda^d$, where  $\lambda^d$ is the diagonal matrix made
with the eigenvalues of $V$. Recall that $A$ is orthogonal if $A^{-1}=A^T$, i.e.
 $\sum_c A_{a c}A_{b c} = \delta_{a b}$.
Then the vector $U^a \equiv A^T X^a$ has the following correlation matrix
$\langle A^T X^a (A^T X^a)^T \rangle = A^T \langle X^a X^{aT} \rangle A =  A^T V A = 
\lambda^d$, i.e. $U^a$ has all its components that are uncorrelated with each other.
The quadratic average $\langle (U^a)^2 \rangle$ of an element of this vector is then 
equal to the corresponding eigenvalue $\lambda_a$. We can thus decompose the assets 
$x^a$ over the set of independent factors $U^a$, in matrix form $X^a = A U_a$ and
explicitely
\be 
x^a_k = \sum_{b=1}^N A_{a b} u^b_k  .
\label{decompp}
\ee
Since the $U^a$ are independent and of variance $\lambda$, the correlation matrix
$\langle X^a X^{aT} \rangle$ is equal to $A \lambda^d A^T$, i.e.
\be
\langle \delta x^a \delta x^b \rangle =  (A \lambda^d A^T)_{ab} = 
\sum_c \lambda_c A_{a c} A_{b c}  . 
\label{correlatt}
\ee
This decomposition thus reduces the problem of correlated assets to the previous case
of uncorrelated assets. Within the Gaussian approximation, the correlation matrix $V$ 
determines completely the distribution of price variations $\delta x^a$ through
\be
P(X^a) \propto \exp \biggl(-{1 \over 2} X^{aT} V^{-1} X^a \biggl) .
\label{distribb}
\ee
We can verify (\ref{defmatcor}) directly from (\ref{distribb}) by making explicit the 
calculation of the correlation with the probability (\ref{distribb}).

\subsection{Optimal portfolio within the mean-variance approach}

Consider a portfolio made of $N$ assets with respective weight $p_a$, with $a=1$  to
$N$. The variation $\delta S(t)$ of the value of the portfolio during the time interval 
$\tau$ is
\be
\delta S(t) = \sum_{a=1}^N p_a \delta x^a(t) 
\ee
($= p^T X^a$ in matrix notation). The variance $\langle [\delta S(t)]^2 \rangle$ is
\be
\langle [\delta S(t)]^2 \rangle = p^T V p ~.
\ee
Notice that we can retrieve this result from (\ref{distribb}) by using the fact that 
the distribution of $\delta S(t)$ is formally $P(\delta S(t))
\propto \int dX^a P(X^a) \delta(S-\sum_{a=1}^N p_a \delta x^a(t))$. Its Fourier transform
can be expressed under the form 
\be
\prod_{i=1}^N (\int dx_i) \exp \biggl(-{1 \over 4} x_i A_{ij}^{-1} x_j + y_i x_i \biggl) =
\sqrt{\pi^N \over det(A_{ij})} \exp (y_i A_{ij} y_j)~,
\label{identt}
\ee
which by inverse Fourier transform gives
\be
P(\delta S(t)) \propto \exp \biggl(-{[\delta S(t)]^2 \over 2 p^T V p}\biggl)  ~.
\label{distriss}
\ee

It is useful to express $\langle [\delta S(t)]^2 \rangle = p^T V p$ under a form that 
exploits the decomposition over the independent components $U^a$. Reporting $X^a = A U^a$ in
$\delta S(t) = p^T X^a$, we obtain $\delta S(t) = {\hat p}^T U^a$, where
\be
{\hat p} \equiv A^T p ~. 
\label{defff}
\ee
This expression represents the portfolio as made of a set of effective assets
$U^a$, $a=1$ to $N$ which are uncorrelated, with effective weights ${\hat p}^a$. We can thus
directly use the previous results as soon as we specify the $U^a$'s. 
We have 
\be
\langle [\delta S(t)]^2 \rangle = {\hat p}^T \lambda^d {\hat p} = \sum_{a=1}^N \lambda_a [{\hat
p}^a]^2  ~. \label{quadrac}
\ee
The average return is given by 
\be
\langle \delta S(t) \rangle= \sum_{a=1}^N p_a \langle \delta x^a(t) \rangle = p^T \langle X^a
\rangle = {\hat p}^T \langle U^a \rangle .
\ee
Within the Gaussian approximation, the risk associated with the portfolio defined in terms of the
weights $p^a$ is quantified by the variance $\langle [\delta S(t)]^2 \rangle -
\langle \delta S(t) \rangle^2$ given from (\ref{quadrac}) by replacing $V$ by the covariance
matrix, i.e. by substracting $\langle \delta x^a \rangle \langle \delta x^b
\rangle$ to $\langle \delta x^a \delta x^b \rangle$). In the sequel, we use the same notation
$V$ for the covariance matrix (and the other derived ones).

The mean-variance approach developed by Markovitz, that we already visited
in the first sections, consists in looking for the weights
$p^a$ such that the return be the largest possible for a given variance
$\langle [\delta S(t)]^2 \rangle - \langle \delta S(t)
\rangle^2$ or equivalently that the variance $\langle [\delta
S(t)]^2 \rangle - \langle \delta S(t) \rangle^2$ be minimum for a given return. 
The solution is obtained by minimizing the following expression with respect to the weights
 $p^a$
\be
M \equiv \sum_{j=1}^N \lambda_j [{\hat p}^j]^2 - \alpha_1 \sum_{j=1}^N {\hat p}^j \langle U^j
\rangle -  \alpha_2 \sum_{j=1}^N {\hat p}^j  ~,
\ee 
where $\alpha_{1,2}$ are the Lagrange parameters introduced to constraint the minimization
at fixed gain $\langle \delta S(t) \rangle$ and to ensure the normalization of the weights.

The extremalization with respect to a weight $p^a$ gives
\be
{\partial M \over \partial p^a} =
\sum_j {\partial M \over \partial {\hat p}^j} {{\partial {\hat p}^j} \over {\partial p^a}} =
\sum_j \biggl( 2 \lambda_j {\hat p}^j - (\alpha_1  \langle U^j \rangle - \alpha_2 ) 
(A^T)_{ja} \biggl)~,
\ee
which gives the following matrix equation 
\be
\lambda^d {\hat p} = \alpha_1 \langle U^a \rangle + \alpha_2 {\vec 1}~,
\ee
where ${\vec 1}$ is the single column vecteur with unit elements.
With (\ref{defff}), $\langle U^a \rangle = A^T \langle X^a \rangle$ and $V = A \lambda^d A^T $, we
finally obtain
\be
p^a = \alpha_1 V_{ab}^{-1}  \langle X^b \rangle + \alpha_2 A_{ab} {\vec 1}^b ~.
\label{solluu}
\ee
Imposing $\sum_{i=1}^N p^i = 1$ and $\sum_{i=1}^N p_i \langle X^i \rangle = 
\langle \delta S(t) \rangle$ yields
$\alpha_1$ and $\alpha_2$ as functions of $\langle \delta S(t) \rangle$. Varying 
$\langle \delta S(t) \rangle$ enables us to derive the mean-variance curve, called the
``efficient frontier''. $\alpha_1$ can be interpreted as the risk aversion parameter.
Numerous books discuss these solutions \cite{portfolio}.

Nothwithstanding a wide application due to its convenience and simplicity, the mean-variance
approach hides several severe problems.
\begin{itemize}
\item Mean-variance portfolios are not very diversified and have the tendency to 
select assets with comparable risks.
\item A frequent reallocation is called for to address the non-stationarity of the
estimation of the covariance matrix.
\end{itemize}
We now expose several successive generalizations of the mean-variance approach that 
specifically address the limitation of the Gaussian approximation.

\subsection{Quasi-Gaussian parametrization}

It may be useful to attempt to represent the multivariate distribution of price variations of
$N$ assets by the following expression
\be
P(X) = F \biggl( (X-X^0)^T V^{-1} (X-X^0) \biggl)  ~.
\label{quasiggaus}
\ee
$X^0$ is the unit column vector of the average of the price variations.
The function $F$ is kept a priori arbitrary. Notice that if $F$ is an exponential, we
retrieve the Gaussian distribution (\ref{distribb}) and $V$ becomes the covariance matrix.
Consider a portfolio with weights given by the unit column vector $p$. Its value variation 
during the unit time is $\delta S(t) = \sum_{a=1}^N p_a \delta x^a(t) = p^T X$ using the matrix
notation. The distribution $P(\delta S) d \delta S$ can be written as
\be
P(\delta S) = \int dX F \biggl( (X-X^0)^T V^{-1} (X-X^0) \biggl)  \delta (\delta S-p^T X) ~.
\label{azetpw}
\ee
To estimate this integral, we isolate one of the assets $x_1 = x^0_1+ y_1$ and write,
using $Y=  X-X^0$, 
\be
Y^T V^{-1} Y = V^{-1}_{11} y_1^2 + 2 (v^T y) y_1 + y^T {\cal V}^{-1} y ~,
\label{froemru}
\ee
where $V^{-1}_{ij}$ is the element $ij$ of the matrix $V^{-1}$, $y$ is the unit column vector
$(y_2, y_3, ...., y_N)^T$ of dimension $N-1$, $v$ is the unit column vector $(V^{-1}_{21},
V^{-1}_{31}, ..., V^{-1}_{N1})^T$ of dimension $N-1$, and ${\cal V}^{-1}$ is the square matrix of
dimension $N-1$ by $N-1$ derived from $V^{-1}$ by removing the first row and first column.
The factor $2$ in $2 (v^T y) y_1$ comes from the symmetric structure of the matrix $V^{-1}$.

We can now express the condition $\delta(\delta S - p^T X)$ in the integral
(\ref{azetpw})\,:
\be
{1 \over p_1}  \delta (y_1 - {1 \over p_1}(\delta S - P^T X_0 - {\cal P}^T y)) ~,
\ee
where  ${\cal P}$ is the unit column vector $(p_2, p_3, ..., p_N)^T$ of dimension $N-1$.
The integration over the variable $y_1$ cancels out the Dirac function and we obtain the
argument of the function $F$ under a quadratic form in the variables $S$ et $y$. Using the
identity
\be
X^T V^{-1} X +  X^T Y = {\hat X}^T V^{-1} {\hat X} - {1 \over 4} Y^T V Y ~,
\ee
where ${\hat X} = X + VY$, we obtain
\be
P(\delta S) = \int d{\hat y} F\biggl( {\hat y}^T M^{-1} {\hat y} + {\delta S^2 \over P^T V P}
\biggl) ~, 
\label{qusrdfgscy}
\ee
where the integral is carried out over the space of vectors ${\hat y}$ of dimension $N-1$ and
\be
M^{-1} \equiv {\cal V}^{-1} - {2 \over p_1} (v - {V^{-1}_{11} \over 2p_1} {\cal P}) {\cal P}^T ~.
\ee
We can finally write
\be
P(\delta S) = {\cal F} \biggl({\delta S^2 \over p^T V p}\biggl) ~,
\label{qusrcy}
\ee
where ${\cal F}(x)$ is defined by (\ref{qusrdfgscy}). The prefactor 
${1 \over P^T V P}$ of $\delta S^2$ is simply deduced by remarking that it is independent of 
the function $f$ and thus equal to that obtained for the Gaussian case.

This remarkable result shows that the typical amplitude of the fluctuations of the 
values of the portfolio is still controlled by the quasi-variance
$p^T V p$ defined as in the Gaussian case (\ref{distriss}). It is then natural
to optimize the portfolio using the quasi-variance as the measure of the risk. This 
parametrization provides a natural generalization of the standard mean-variance Markovitz
approach. There is however a danger for the largest fluctuations in relying only on this
insight. Indeed, the expression (\ref{qusrdfgscy}) for ${\cal F}(x)$ shows that the 
explicit dependence of the distribution $P(\delta S)$ is in general a function of the
asset weights constituting the portfolio, beyond the simple dependence in $p^T V p$. 
Minimizing only $p^T V p$ may thus be insufficient because it may be linked to a 
dangerous deformation of ${\cal F}(x)$ in the tail.

\subsection{Generalization to non-gaussian correlated assets}

We assume that it is still possible to decompose the assets
$X^a$ on a set of effective independent assets $U^a$\,:
\be
X^a = A U^a ~. 
\label{decomp}
\ee
The unit column vector $U^a$ can be interpreted as the set of ``explanatory'' factors of
the price variations. For this decomposition to be useful, the matrix $A$ should be constant
while $X^a$ and $U^a$ fluctuate in time. This provides a set of orthogonal assets whose
fluctuations are uncorrelated. An estimation of the matrix $A$ can be obtained from the 
covariance matrix as above. A first generalization consists in relaxing the condition that 
the $U^a$ be distributed according to a Gaussian distribution as in the previous section.
From the data $X^a$ and the construction of the matrix $A$, we can study each effective 
asset $U^a$ and determine its distribution $P_a^U(U^a)$. If the decomposition works and
the effective assets are not correlated, we can write for each effective asset $U^a$
the cumulant expansion of the Laplace transform of $P_a^U(U^a)$ under the form (\ref{cumulantt}).
Since the variation of the portfolio value is given as before by 
$\delta S(t) = p^T X^a = {\hat p}^T U^a$, we obtain the Cram\'er function of the
distribution of $\delta S(t)$ under the form (\ref{cumulbb}) with (\ref{ccc},\ref{cccc})
where the $p_j$'s are replaced by ${\hat p}^j= A^T p$ according to the equation
(\ref{defff}). The cumulants corresponds to the effective assets $U^a$. We thus obtain
the distribution of $\delta S(t)$ which accounts for the first leptokurtic corrections
given by (\ref{gaussccc}) for the non-symmetric case and by
(\ref{gaussddd}) for the symmetric case.

Let us consider the symmetric case (\ref{gaussddd}). This expression shows that the risk
is not uniquely represented by the variance $V$ but also by the coefficient $C_4$ as well as
by all higher order cumulants. There is a danger in working only on the variance because minimizing
only the variance may lead to larger fluctuations than before the minimization! To see this, 
we note that typical fluctuations $S-\langle S \rangle \sim V$ have a probability becoming
much larger than the Gaussian estimate because, if $V$ becomes small 
by the action of the naive mean-variance optimization,
when $C_4$ does not decrease in proportion, the correction term ${C_4 \over V}$ 
remains large. At this order in the perturbative expansion, it is 
necessary to take into account non only $V$ but also $C_4$. The minimization of $V$
must be performed by controlling and even maximizing simultaneously the ratio
${V^2 \over C_4}$, this at fixed return. The second condition ensures that the weight of the 
non-Gaussian tail remains small, thus mastering the large risks. 

This optimization problem has no unique solution due to the existence of several 
contradictory constraints. In standard economic theory, one relies on the 
use of utility function to decide the relative importance of the different constraints.
For the sake of illustration, let us propose the following strategies.
\begin{itemize}
\item Minimize $V$ while maximizing ${V^2 \over C_4}$ can be obtained by minimizing the ratio
\be
V/({V^2 \over C_4})^{\beta} = V^{1-2\beta}~C_4^{\beta}~~~~~~{\rm with}~~~0 \leq \beta < {1 \over 2}~,
\ee
to ensure that both $V$ and $C_4$ are simultaneously decreased.
$\beta$ quantifies the relative weight given to the kurtosis.

\item One can introduce a parameter $\psi$ which quantifies the sensitivity of the investor 
with respect to large fluctuations measured by $C_4$ and minimize the function
\be
V - \psi {V^2 \over C_4} - \alpha_1 R - \alpha_2 \sum_j p_j ~,
\ee
where $\alpha_1$ ensures a fixed return and 
$\alpha_2$ accounts for the normalization of the asset weights in the portfolio.
\end{itemize}

\subsection{Exponential distributions}

We have already analyzed the problem of a portfolio constituted of assets with exponentially
distributed price variations. Let us reexamine this problem
within the framework of the cumulant expansion. We work directly on the effective explanatory
assets $U^a$ that are non-correlated.

The distribution $P_S(\delta S)$ is given by (\ref{produit}) with the 
$p_j$'s replaced by ${\hat p}_j$ (in the sequel, we omit the hat). The
coefficients $\alpha_j$ are the exponents of the exponential distributions of price 
variations of the independent assets. An expansion in power of $k^2$ as in (\ref{devel}) but 
up to the order $k^4$ gives
\be
P_S(\delta S) =  {1 \over 2\pi} \int_{-\infty}^{+\infty} dk e^{ik \delta S} 
\exp \biggl( - k^2 \sum_{j=1}^N (p_j\alpha_j^{-1})^2  + {k^4 \over 2} 
\sum_{j=1}^N  (p_j\alpha_j^{-1})^4 + ...   \biggl) ~.
\ee
$P_S(\delta S)$ is thus of the form (\ref{gaussccc}) with
$V=2 \sum_{j=1}^N  (p_j \alpha_j^{-1})^2$ and $C_4 = \sum_{j=1}^N (p_j\alpha_j^{-1})^4$.
Let us assume that all assets have the same return, fixed to zero without loss of generality.
We thus focus our analysis to the minimization of the risk.

The first scenario consists according to the Gaussian approach to minimizing the variance
${{\sum_{j=1}^N (p_j\alpha_j^{-1})^2} \over {(\sum_{j=1}^N p_j)^2}}$. The solution is
\be 
p_k^{(1)} = {\alpha_k^2 \over \sum_{j=1}^N \alpha_j^2} ~ . 
\label{premstrat}
\ee

The second scenario consists in focusing on the large risks, here quantified by the ratio
${V^2 \over C_4}$ which describes the large fluctuations of the price variations.
The ratio ${V^2 \over C_4}$ can be written as ${(\sum_{j=1}^N W_j)^2
\over  \sum_{j=1}^N W_i^2}$ with $W_i \equiv (p_j \alpha_j^{-1})^2$. It is maximum when all
the weights $W_i$ are equal, i.e.
\be 
p_k^{(2)} = {\alpha_k \over \sum_{j=1}^N \alpha_j} ~. 
\label{seconstrat}
\ee
This is the result (\ref{resythgfglk})  already obtained from the condition of minimizing 
the probability of large losses.

Notice that the two strategies (\ref{premstrat}) and (\ref{seconstrat}) become identical 
$p_k= 1/N$, if all assets have the same parameter $\alpha$. Indeed, the large risks are
diversified from the beginning since the probability tails are identical in this case.
The strategy (\ref{seconstrat}) corresponds to balance the risks equally over all assets.
In contrast, the portfolio (\ref{premstrat}) over-controls the balance\,: if an asset exhibits
a larger risk in the tail, i.e. its coefficient $\alpha$ is significantly smaller than
the others, it is almost completely absent from the portfolio. If an asset has a large 
$\alpha$, it will have a large weight in the portfolio and its parameter 
${\alpha \over p}$ becomes smaller than the other ones. We see here the general trend
exhibited by solutions obtained from the mean-variance approach which correspond to a 
diversification only on assets presenting comparable risks, while excluding almost
completely the more risky assets.

The strategy (\ref{seconstrat}), which implies that $p_k^{(2)}\alpha_k^{-1} = \sum_{j=1}^N
\alpha_j$ is a constant independent of $k$, leads to a simple exact expression of the 
distribution of the fluctuations $\delta S$ of the portfolio value. Indeed, the
formula (\ref{produit}) gives in a manner similar to (\ref{exactt})\,:
\be
P_S(\delta S) = {\sum_{j=1}^N \alpha_j \over \Gamma(N)} 
\biggl([\sum_{j=1}^N \alpha_j] \delta S \biggl)^{N-1} 
\exp \biggl(-[\sum_{j=1}^N \alpha_j] \delta S\biggl) ~ .
\label{exacttt}
\ee
This Gamma distribution converges in its center to the Gaussian law while keeping 
an exponential tail for the largest fluctuations, with an exponent $\sum_{j=1}^N \alpha_j$.
This can be compared to the distribution (\ref{resul}) with the value of $p_j$'s given by 
(\ref{premstrat})\,:
\be 
P_S(\delta S) = {1 \over 4} \sum_{j=1}^N {\sum_{i=1}^N  \alpha_i^2 \over \alpha_j}
{1 \over \prod_{i \neq j} \biggl(({\alpha_j \over \alpha_i})^2 - 1 \biggl)}
\exp \biggl(-{\sum_{i=1}^N  \alpha_i^2 \over \alpha_j} |\delta S|\biggl)  ~.
\label{resull}
\ee
On see clearly that, as expected, the large losses are better controlled by the 
strategy (\ref{seconstrat}) yielding typical fluctuations of the order of
$\delta S^{(2)} \sim  [\sum_{j=1}^N \alpha_j]^{-1}$, while the strategy (\ref{premstrat}) 
leads to typical fluctuations of order $\delta S^{(1)} \sim {\alpha_{max} \over \sum_{i=1}^N 
\alpha_i^2}$, where  $\alpha_{max}$ is the largest of the exponents. Let us assume for the sake
of illustration that the exponents $\alpha$ are distributed according to a Gaussian law with mean
$\langle \alpha \rangle$ and variance $\sigma^2$, then 
$\delta S^{(2)} \sim {\langle \alpha \rangle^{-1} \over N}$ and
$\delta S^{(1)} \sim {\langle \alpha \rangle^{-1} \over N} + {\sigma \over 
\langle \alpha \rangle^2} {\sqrt{\log N} \over N}$.

\subsection{General theory and cumulant expansion from the generalized correlation matrices}

\subsubsection{Strategies}

Let us consider the case where the assets are distributed according to the 
arbitrary joint distribution
\be
P(x_1,x_2, ...,x_i,...,x_N, t/ x_1^0,x_2^0, ...,x_i^0,...,x_N^0, 0)~,
\ee
which is in general non-Gaussian and with arbitrary inter-asset correlations.
A natural strategy to determine a portfolio is to calculate the weights
$p_k$ of the assets in the portfolio, $k=1$ to $N$, which minimize the risk of losses
larger than a chosen VaR $\lambda$, this in the presence of other constraints for instance on the
return\,:
\be
{\partial \biggl(\int_{-\infty}^{-\lambda} P_S(\delta S) d\delta S - \alpha_1 \sum_{j=1}^N p_j 
\langle \delta x_j \rangle - \alpha_2 \sum_{j=1}^N p_j \biggl)  \over \partial p_k} = 0 
\label{condddd}
\ee
for $k=1$ to $N$, where $\alpha_1$ and $\alpha_2$ are Lagrange parameters.
 $\lambda$ becomes a parameter that quantifies the risk aversion of the investor.
 The residual probability ${\cal P}_{min}(\lambda)$ obtained after the optimization gives
 the mean frequency of the occurrence of losses equal or 
 larger than $\lambda$. This approach requires a three-dimensional representation along the
 following axis\,:
 \begin{enumerate}
\item the VaR $\lambda$ controlling the acceptable level of loss,
\item the loss probability ${\cal P}(\lambda)$ at this level,
\item the expected return.
\end{enumerate}
It generalizes the usual two-dimensional representation of the mean-variance Markovitz diagram.
This is the price to pay when the distributions of asset price variations are not 
quantified by a single risk parameter, the variance. Note that there is another situation
where the risks can be captured by a single parameter, i.e. the case of power law distributions
quantified by the scale parameter. This case has been discussed previously \cite{nousport}
and the general solution for the portfolio optimization has been given \cite{nousport} as a 
rather straightforward generalization of  the usual mean-variance approach.

In the same spirit as for the minimization of the probability of losses larger than VaR,
let us mention the strategy consisting in maximizing the probability to obtain a minimum
return $\int_{\delta S_{min}}^{\infty} P_S(\delta S) d\delta S$
\cite{Tsai}. Generalizing further, we can propose to minimize the weight
$\int_{-\infty}^{-\lambda} P_S(\delta S) d\delta S$ of the losses in the presence of the 
constraint that the probability of a minimum gain is fixed, thus generalizing in the 
probability space the mean-variance concept. This view point is stimulated by the 
observation that the returns and losses aggregated over long period of times are often
mainly caused by large amplitude price variations that occurred over a very tiny fraction of 
the total time of the investment. For instance, for the US 
$S\&P500$ index, from $1983$ to $1992$, $80 \%$ of the total return stems from 
$1.6 \%$ of the trading time. This leads to optimization problems similar to (\ref{condddd}).

The most general approach consists in first determining the complete distribution 
$P(\delta S(t))$ of the price variations of the portfolio as a function of the distributions
of the underlying assets. Once characterized, one knows fully the impact of the asset weights
on the portfolio. The optimization of the portfolio can then proceed using 
suitable risk measures.

\subsubsection{Distribution of the price variations of the portfolio}

$P(\delta S(t))$ is formally given by
\be
P(\delta S(t)) = \int dX^a P(X^a) \delta(S-\sum_{a=1}^N p_a \delta x^a(t)) ~ ,
\label{dis}
\ee 
where we use again the matrice notation
\be
P(X^a) = P(\delta x_1,\delta x_2, ..., \delta x_i,... \delta x_N)
\ee
and $\int dX^a$ stands for $N$ integrals over the price variations of the $N$ assets.
The probability being positive, we can parametrize it without loss of generality
\be
P(X^a) \equiv \exp \biggl( - {1 \over 2}  X^{aT} V^{-1} X^a - {\cal V}(X^a) \biggl) ,
\label{potentiell}
\ee
where we distinguish a Gaussian part $\exp [ - {1 \over 2}  X^{aT} V^{-1} X^a]$ and a residual
term $\exp [ - {\cal V}(X^a) ] $  representing the non-Gaussian part. It may be dominant over
the Gaussian part. The expression (\ref{potentiell}) defines the Gaussian correlation matrix
$V$ and the non-Gaussian contribution ${\cal V}$, which contains all terms with power 
strickly larger than  $3$ in $X^a$.
With (\ref{dis}), we obtain the expression of the Fourier transform of $P(\delta S)$\,:  
\be   
{\hat P}(k) = 
\int dX^a \exp \biggl( - {1 \over 2}  X^{aT} V^{-1} X^a - {\cal V}(X^a) + H^T X^a \biggl) ~,
\label{laplb}
\ee
where $H \equiv ikp$ and $p$ is as before the unit column vector of the asset weights in the
portfolio.

The most general and powerful technique to determine $P(\delta S)$ from
the exact expression (\ref{laplb}) consists in using the ``technology'' of functional
integrals \cite{Brezin}. The basic idea is to reduce the evaluation of the integrals to 
that of Gaussian integrals. The key technical remark is
$$
{\partial  \over \partial H^c}
\int dX^a \exp \biggl( - {1 \over 2}  X^{aT} V^{-1} X^a - {\cal V}(X^a) + H^T X^a 
\biggl)  =
$$
\be
\int dX^a  X^c \exp \biggl( - {1 \over 2}  X^{aT} V^{-1} X^a - {\cal V}(X^a) + H^T X^a 
\biggl) ~.
\label{ident}
\ee
Expanding the exponentiel as a formal series, we use the identity (\ref{ident}) and then
resum the series to obtain formally
\be
{\hat P}(k) = \exp \biggl(-{\cal V}({\partial  \over \partial
H^c})\biggl)
\int dX^a  \exp \biggl( - {1 \over 2}  X^{aT} V^{-1} X^a + H^T X^a \biggl) ~.
\label{identat}
\ee
The notation ${\cal V}({\partial  \over \partial H^c})$ means that each component
$X^c$ is replaced by the operator ${\partial  \over \partial H^c}$. By this trick,
we have transformed a non-Gaussian integral into an operator applied on a Gaussian
integral that can be calculated explicitely by using
(\ref{identt}). This yields
\be
{\hat P}(k) = \exp \biggl(\sum_{c=1}^N {\cal V}({\partial  \over \partial H^c})\biggl) 
\exp \biggl( {1 \over 2}  H^{aT} V H^a  \biggl) .
\label{identata}
\ee
Notice that we retrieve the Gaussian case for ${\cal V}=0$, i.e. the Fourier transform of
(\ref{distriss}) by replacing $H$ by $ik p$. By the normalization of probabilities, 
${\hat P}(0) = 1$.

One can show \cite{Brezin} that ${\hat P}(k)$ given by
(\ref{identata}) can be written as
\be
{\hat P}(k) \equiv e^{W(H)} ~ ,
\label{WWW}
\ee
where $W(H)$ presents a systematic expansion
\be
W(H) = \sum_{n=1}^{\infty} {1 \over n!} \sum_{j_1=1}^N ... \sum_{j_n=1}^N  H_{j_1} ... H_{j_n}
G_c^{(n)}(j_1,...,j_n) ~ . 
\label{GGG}
\ee
$H_j$ for $j=1,...,N$ is one of the components of the unit column vector $ikp$. The functions
$G_c^{(n)}(j_1,...,j_n)$ can be expressed explicitely in terms of ${\cal V}$
\cite{Brezin}. The usefulness of this formulation is that each $H$ brings in a power of $k$\,:
\be
{\hat P}(k) = \exp \biggl( 
\sum_{n=1}^{\infty} {(ik)^n \over n!} \sum_{j_1=1}^N ... \sum_{j_n=1}^N  
p_{j_1} ... p_{j_n} G_c^{(i)}(j_1,...,j_n)  \biggl)  ~.
\label{explll}
\ee
This expansion defines the cumulants $c_n \equiv (-i)^n {d^n \over dk^n} {\hat
P}(k)|_k=0$ of the distribution $P(\delta S)$\,:
\be
c_n = \sum_{j_1=1}^N ... \sum_{j_n=1}^N p_{j_1} ... p_{j_n} G_c^{(i)}(j_1,...,j_n) ~.
\label{cumufdqsdfqqays}
\ee
Notice that the correlations between the different assets are taken into account
in the functions $G_c^{(i)}(j_1,...,j_i)$. They play the role of 
generalized correlation functions which quantify the pairwise, triplets, etc, correlations
(a multiplet can contain the same asset several times, thus capturing the effect of
self-correlations).

One can also use another systematic expansion
\be
{\hat P}(k) = 
\sum_{n=1}^{\infty} {(ik)^n \over n!} \sum_{j_1=1}^N ... \sum_{j_n=1}^N  
p_{j_1} ... p_{j_n} G^{(i)}(j_1,...,j_n)  ~ ,
\label{expll}
\ee
where the functions $G^{(i)}(j_1,...,j_n)$ can be expressed in terms of the
$G_c^{(i)}(j_1,...,j_n)$. The functions 
$G^{(i)}(j_1,...,j_n)$ are analogous to the moments which can be related to 
$G_c^{(i)}(j_1,...,j_n)$ which are analogous to the cumulants.

Let us persue this analysis and consider the symmetric case where the first term in the
expansion of ${\cal V}(X)$ is quartic\,:
$$
{\cal V}(X) = \sum_{a=1}^N \sum_{b=1}^N \sum_{c=1}^N \sum_{d=1}^N  
v_{abcd} X^a X^b X^c X^d +
$$
\be
 \sum_{a=1}^N \sum_{b=1}^N \sum_{c=1}^N \sum_{d=1}^N  \sum_{e=1}^N
\sum_{f=1}^N   v_{abcdef} X^a X^b X^c X^d X^e X^f + ....
\label{devfyt}
\ee
Keeping for the time being only the quartic terms proportional to $v_{abcd}$, we obtain
$$
{\cal V}({\partial  \over \partial H^c}) \exp \biggl( {1 \over 2}  H^{aT} V H^a  \biggl) =
\sum_{a=1}^N \sum_{b=1}^N \sum_{c=1}^N \sum_{d=1}^N 
v_{abcd} \biggl( V_{ab} V_{cd} +  V_{ac} V_{bd} +  V_{ad} V_{bc} +
$$
\be
V_{ab} S_c S_d + V_{ac} S_b S_d + V_{ad} S_b S_c + V_{bc} S_a S_d + V_{bd} S_a S_c +
V_{cd} S_a S_c + S_a S_b S_c S_d \biggl) ~,
\label{azxvxwhfk}
\ee
where $S_a \equiv ik \sum_{k=1}^N V_{ak} p_k$. We thus see all possible term combinations.
In order to retrieve the fourth order cumulant, we notice that
$$
\exp \biggl[ v_{abcd} \biggl( V_{ab} V_{cd} +  V_{ac} V_{bd} +  V_{ad} V_{bc} +
$$
$$
V_{ab} S_c S_d + V_{ac} S_b S_d + V_{ad} S_b S_c + V_{bc} S_a S_d + V_{bd} S_a S_c +
V_{cd} S_a S_c + 
$$
\be
(1 - {1 \over 2} (V_{ab}V_{cd} + V_{ac}V_{bd} + V_{ad}V_{bc}))
S_a S_b S_c S_d \biggl) \biggl]
\ee
retrieves (\ref{azxvxwhfk}) by an expansion to the quartic order. Here, we simply use the
fact that $\exp (ax + bx^2) = 1 + ax + ({1 \over 2} a^2 +
b) x^2 + ...$. We thus obtain 
\be
c_4 = \sum_{j=1}^N \sum_{k=1}^N \sum_{l=1}^N \sum_{m=1}^N  G_c^4(j,k,l,m) p_j p_k p_l
p_m ~, \label{cumufays}
\ee
with
\be
G_c^4(j,k,l,m) = 24 \sum_{a=1}^N \sum_{b=1}^N \sum_{c=1}^N \sum_{d=1}^N 
v_{abcd} \biggl( 1 - {1 \over 2} (V_{ab}V_{cd} + V_{ac}V_{bd} + V_{ad}V_{bc}) \biggl)
V_{aj}V_{bk} V_{cl} V_{dm} ~. 
\label{cuytrwx}
\ee
The higher order terms like $v_{abcdef} X^a X^b X^c X^d X^e X^f$ also contribute to the fourth
order cumulant as seen by generalizing (\ref{azxvxwhfk}). In fact, 
$c_4$ receives contributions from all higher order terms. They are weighted by the 
coefficients $v_{abcdef...}$ which in general decrease rather fast. Diagramatic techniques
can then be used to keep track of all terms at a given order in the systematic expansion
 \cite{Brezin}.

\subsubsection{Application to the quasi-gaussian case}

The quasi-gaussian case where the distribution $P(\delta S)$ has the form
(\ref{qusrcy}) implies precise constraints on the structure of the cumulants of 
$P(\delta S)$ and thus on the correlation functions between the assets. Indeed, from the 
expression (\ref{gaussddd}) giving $P(\delta S)$ up to the first correction in terms of the 
kurtosis, we see that $P(\delta S)$ is uniquely a function of ${\delta S^2 \over P^T V P}$
(where $P^T V P$ is 
denoted $V$ in (\ref{gaussddd})) only if the cumulant of order $4$ is proportional to $V^2$
with a coefficient of proportionality which is a pure number. As a consequence, the kurtosis
must be a number independent of the asset weights in the portfolio. For this to be true, the
cumulant $c_4$ given by (\ref{cumufays}) must factorize and is proportional to the square of the
cumulant $c_2$\,:
\be 
c_2 \equiv \sum_{j=1}^N \sum_{k=1}^N  G_c^2(j,k) p_j p_k ~, 
\label{cuermufays}
\ee
where
\be
G_c^2(j,k) =  \sum_{a=1}^N \sum_{b=1}^N V_{aj}V_{bk} ~.
\label{cumureysf}
\ee
The identification term by term yields
\be
G_c^4(j,k,l,m)  =  w G_c^2(j,k)  G_c^2(l,m) ~,
\label{conffytyrz}
\ee
where $w$ is arbitrary. This expression (\ref{conffytyrz}) together with
(\ref{cuytrwx}) and (\ref{cumureysf}) determines the particular structure of the 
four-asset correlations $v_{abcd}$ in the quasi-gaussian case (\ref{qusrcy}).

\section{Conclusion}

We have tried to demonstrate the analogies between the quantification of risks in 
finance and insurance and the optimization of portfolios on one hand and statistical physics
concepts and methods on the other hands. The main message similar to that given long ago for random 
physical systems \cite{Anderson} is that a suitable risk assessment requires the study of the
full distributions of price variations in contrast to the more standard variance approach.
We have also shown how tools developed in statistical physics to address large fluctuations
can be used in the optimization of portfolios. We thus hope to foster the interest of
the physical community in these fascinating problems.

\vfill\eject

\pagebreak

FIGURE CAPTIONS

\vskip 1cm

Figure 1: Return $r$ and variance $v$ as a function of $f$ for $p=1/2$ and $\lambda = 2.1$.

\vskip 1cm

Figure 2: Return $r$ as a function of $v$ for the same values $p=1/2$ and $\lambda = 2.1$
as in figure 1.

\vskip 1cm

Figure 3a: The characteristic time $t^*(f)$ defined by
(\ref{tempscharafgf}) as a function of $f$ for $r \geq 0$.

\vskip 1cm

Figure 3b: The Sharpe parameter $Sharpe_1 = \sqrt{2} [t^*(f)]^{-1/2} = {r \over \sqrt{v}}$
defined for a unit time step as a function of $f$.

\vskip 1cm

Figure 4: Dependence of the average wealth $\langle S(1) \rangle$ and of the 
typical wealth $S_{pp}(1)$ as a function of $f$ after one time step, for the case
 $p=1/2$ and $\lambda = 2.1$. 
 
 \vskip 1cm

Figure 5a: The Cram\'er function $s(x)$ given by (\ref{cramerc}) and its parabolic 
approximation as a function of $x$ for $p=0.5$.

\vskip 1cm

Figure 5b: The Cram\'er function $s(x)$ given by (\ref{cramerc}) and its parabolic 
approximation as a function of $x$ for $p=0.95$.

\vskip 1cm
Figure 6: The Cram\'er function
$s(y \equiv \alpha \delta S) = \log (y) + 1 - y$ and its
parabolic approximation $s_g(y) = -{1 \over 2} (y - 1)^2$. 

 \end{document}